\documentclass{sig-alternate-05-2015}

\usepackage{lscape}
\usepackage{fixltx2e}
\usepackage{graphicx}
\usepackage{booktabs}

\usepackage{epigraph}
\newtheorem{Definition}{Definition}

\def\pf{\hbox{\textit{p}\hskip-1pt\emph{\textsubscript{F}}}}
\def\pfn{\hbox{\textit{p}\hskip-1pt\emph{\textsubscript{F\hskip-1pt N}}}}
\def\pfp{\hbox{\textit{p}\hskip-1pt\emph{\textsubscript{F\hskip-1pt P}}}}
\def\totalfp{\hbox{$\sum{\!\raise-0.25ex\hbox{\emph{F\hskip-1pt P}}}$}}
\def\totalfn{\hbox{$\sum{\!\raise-0.25ex\hbox{\emph{F\hskip-1pt N}}}$}}
\def\totaltn{\hbox{$\sum{\!\raise-0.25ex\hbox{\emph{T\hskip-1pt N}}}$}}
\def\totaltp{\hbox{$\sum{\!\raise-0.25ex\hbox{\emph{T\hskip-1pt P}}}$}}
\def\totalcost{\mbox{$\sum{\!\raise-0.25ex\hbox{\textup{\textit{SC}}}}$}}
\def\totaltime{\mbox{$\sum{\!\raise-0.25ex\hbox{\textup{\textit{WT}}}}$}}

\def\FNA{\hbox{$F\hskip-1pt N_{\hskip-1pt A}$}}
\def\FNG{\hbox{$F\hskip-1pt N_{G}$}}
\def\FPA{\hbox{$F\hskip-1pt P_{\hskip-1pt A}$}}
\def\FPG{\hbox{$F\hskip-1pt P_{G}$}}

\usepackage{scalerel}

\newcommand\reallywidehat[1]{\arraycolsep=0pt\relax%
\begin{array}{c}
\stretchto{
  \scaleto{
    \scalerel*[\widthof{\ensuremath{#1}}]{\kern-.5pt\bigwedge\kern-.5pt}
    {\rule[-\textheight/2]{1ex}{\textheight}} 
  }{\textheight} %
}{0.5ex}\\           
#1\\                 
\rule{-1ex}{0ex}
\end{array}
}

\def\NCSC{\hbox{$\widehat{\totalcost}$}}
\def\NTT{\hbox{$\widehat{\totaltime}$}}
\def\NTC{\hbox{$\pmb{\sharp}$}}

\begin{document}
\conferenceinfo{iiWAS}{11--13 Dec. 2015, Brussels, Belgium}
\crdata{978-1-4503-3491-4}

\title{Tapping Into the Wells of Social Energy:\\
       A Case Study Based on Falls Identification}
\numberofauthors{2}
\author{
%
%
\alignauthor
Vincenzo De~Florio\\
       \affaddr{MOSAIC/Universiteit Antwerpen \&}\\
       \affaddr{MOSAIC/iMinds research institute}\\
       \affaddr{Middelheimlaan 1, 2020 Antwerp, Belgium}\\
       \email{Vincenzo.DeFlorio@uantwerpen.be}
\alignauthor
Arianit Pajaziti\\
       \affaddr{MOSAIC/Universiteit Antwerpen}\\
       \affaddr{Middelheimlaan 1, 2020 Antwerp, Belgium}\\
       \email{Arianit.Pajaziti@student.uantwerpen.be}
}
%


\maketitle

\begin{abstract}
Are purely technological solutions the best answer we can get to the shortcomings
our organizations are often experiencing today? The results we gathered in this work lead us
to giving a negative answer to such question. Science and technology are powerful
boosters, though when they are applied
to the ``local, static organization of an obsolete yesterday'' they fail to translate in the
solutions we need to our problems. Our stance here is that those boosters should be
applied to novel, distributed, and dynamic models able to allow us to escape from
the local minima our societies are currently locked in. One such model is simulated
in this paper to demonstrate how it may be possible to tap into the vast basins of social energy
of our human societies to realize ubiquitous computing sociotechnical services for
the identification and timely response to falls.
\end{abstract}

\ccsdesc[500]{Applied computing~Health care information systems}
\ccsdesc[500]{Computing methodologies~Simulation evaluation}
\ccsdesc[500]{Human-centered computing~Collaborative and social computing systems and tools}

\printccsdesc


\terms{Theory---change this with the real term}

\keywords{Falls identification; telecare; sociotechnical systems; agent-based simulation; role-flow}

\section{Introduction}
\epigraph{I suppose it is tempting,\\
if the only tool you have is a hammer,\\
to treat everything as if it were a nail.}%
{{Abraham Maslow,}\\ \emph{The Psychology of Science}}

A well-known syndrome introduced by Abraham Maslow is the so-called ``law of the instrument'', stating that
solutions to problems are often influenced by the tools available off-the-shelf~\cite{Maslow}.
This obviously introduces
limitations and inefficiencies. A typical case in point is given by information and communication technology
(ICT) solutions to the problem of fall identification. Falls are the ``most significant cause of injury for elderly
persons''~\cite{EDC2013}---more specifically ``the most serious life-threatening events that can occur'' in the
65+ age group~\cite{Benda13}. Although not a frequent event, it is one whose consequences are often of the most
concern~\cite{Benda13,Cusano14}, especially in the case proper treatment is not timely dispatched to the
fallen person. A common ``tool'' to deal with this problem is fall detection devices. Such devices---for instance
accelerometers, gyroscopes, or other sensors---constantly monitor a person and trigger an alarm
when some safety threshold is reached. A major limitation and inefficiency of this approach is due to the
current difficulties of our algorithms and their implementations
to provide reliable assessments of real-life situations.
Despite the continuing technological progress, it is still very difficult for an ICT
monitoring device to determine whether a person has actually fallen or if, e.g., he or she has knelt
down very quickly.
The use of redundancy---by coupling, for instance, two accelerometers of different technology and design)
improves the sensitivity (i.e., reduces the false negative rate), though this is often reached
at the price of increasing the false alarm ratio (i.e., reducing the specificity). The already high
social costs are thus exacerbated by unnecessary interventions due to false alarms.

How to deal with this problem? One way is to try and improve the current tools. Purely ICT solutions to fall
detection assessment are the subject of numerous research actions, which results in improvements in both
sensitivity and specificity. Despite those efforts, ``no silver bullet [is] in sight [yet]''~\cite{Cusano14}.
Another possibility is to widen our horizons and look for different and better approaches.
In this paper we discuss one such approach.
We propose to go beyond the purely technological solution and to integrate humans in the system.
The resulting socio-technological system is introduced and analyzed here by means of a multi-agent
simulation model. We show that by integrating information from the cyber- and the physical
domain and by appointing verification tasks to a ``cloud'' of human agents
it is possible to overcome, to some extent, the current limitations and inefficiencies of purely ICT-based
solutions.

In what follows we first briefly recall in Sect.~\ref{s:def} a number of definitions.
Section~\ref{s:fso} then briefly summarizes the key elements of our fractal social organizations---a
distributed, bio-inspired organization that we have used to structure the simulation models employed
in this paper.
Section~\ref{s:act} follows and introduces our actor classes.
Two classes of simulation models corresponding to
the use of either one or two fall detectors coupled with a variable
number of human agents are given in Sect.~\ref{s:sce}.
Results are discussed in Sect.~\ref{s:res}, while
Sect.~\ref{s:end} recalls the major lessons learned and concludes.

\typeout{As Falls Wichita, so Falls Wichita Falls}

%
%

\section{Figures of Interest}\label{s:def}
We now introduce the figures we focus our attention on in the course of our simulations.

\begin{Definition}[False positive rate]
False Positive\\
\emph{(FP)} rate (commonly known also as
False Alarm rate) is the probability of concluding that the observed facts do imply the
identification of a situation, when in fact that is not the case.
The event did not take place, but the system ``fired''.
Every occurrence of said wrong conclusion is called a \emph{FP}.
\end{Definition}

In the face of a FP, assistance is delivered but results in a waste of resources.
Social costs go up without providing any social returns.

\begin{Definition}[Specificity]
Specificity is the probability of correctly identifying that a given event has not taken place.
It is equal to $1 - \hbox{\emph{FP} rate}$.
\end{Definition}

\begin{Definition}[False negative rate] 
False negative \emph{(FN)} rate is the probability of not being able to identify the occurrence
of an event that took place.
The event was experienced, though the system did not ``fire''.
Every time the system fails to identify a true fall we shall say that
a \emph{FN} occurred. In other words, a \emph{FN}
is a missed alarm: the system did not react in the face of a true fall.
\end{Definition}

\begin{Definition}[Sensitivity]
Sensitivity is the probability of correctly identifying that a given event has taken place.
Sensitivity is equal to $1 - \hbox{\emph{FN} rate}$.
\end{Definition}

As efficaciously observed by Dr. Tom Doris, KeepUs project founder and technical lead~\cite{Cusano14},
\begin{quote}
``In any safety system, false negatives are possibly the worst kind of failure.''
\end{quote}
This is particularly true in the case of falls. Quoting again Dr. Doris,
\begin{quote}
``the single most important factor influencing the long-term outcome [after a fall]
is the length of time between the fall and getting medical attention at a hospital.
A few hours more or less makes the difference between life and death.''
\end{quote}
Evidence of a strict correlation between
the time of arrival
of medical caregivers and the mortality rate may be found also in~\cite{Benda13,Bourke10}.

A problem we tackle in this paper is the well-known correlation between
FP and FN. Whatever the method in use today,
if one wants to reduce FN rate, usually FP rate goes up~\cite{Cusano14}.
In what follows we are going to introduce a method to deal with this problem
based on the use of ``social energy,'' which we
defined in~\cite{DeBl10} as ``the self-serve, self-organization, and self-adaptability potentials of our societies.''

As a first component of our solution we now briefly introduce
the main elements of the organizational structure that we call
``fractal social organizations.''

\section{Fractal Social Organizations}\label{s:fso}
Fractal Social Organizations (FSO)
is an organizational structure that realizes a
nested compositional hierarchy~\cite{HT:TE14a} (NCH): the system is a network
of nodes, each of which is a network of other nodes.
FSO nodes are called circles. Being a NCH means that the
FSO is structured as concentric circles. Examples of said circles may be for instance
an employee of a business enterprise; an office of that enterprise; the whole business enterprise;
or a business ecosystem including several business bodies.
In each FSO circle there is a node that coordinates that circle.
Said coordinator node receives notification data
from all the nodes of its circle.
Notifications received by the coordinator describe the evolving of situational and context
information. This means that any node state change and also any detected external
change are reported\footnote{In our prototypical implementation of the FSO
this was implemented via a standardized, asynchronous
publish-and-subscribe mechanism~\cite{Oasis2}.}.

The circle coordinator is an autonomic computing agent implementing a MAPE loop. Received
data is fed into the loop's ``M'' component---namely, its perception organ. Once received, data is analyzed
and integrated (by the ``A'' component) and semantically matched with the information already known to the coordinator.
Deductions are fed into a
the planner component (``P''), which selects a response protocol through a simple algorithm.
The selected protocol is then executed by component ``E''.


Protocols are associated at design time to state transitions and situation identifications.
If, for instance, situation $i$ = ``fall suspected in flat 145'' is associated to protocol $p_i$, then
as soon as $i$ is identified protocol $p_i$ is readied for running.

A readied protocol is a protocol that has been authorized for launching though is not running yet.
Protocols in fact require the availability of one or more \emph{agents}. Protocol $p_i$
may need, e.g., the cooperation of a general practitioner, a nurse, an ambulance, an ambulance driver,
and specific medical devices.

In order to launch a protocol, the coordinator needs to enroll agents: assign agents to the needed roles.
This is similar to data-flow, only it requires active roles instead of data. Because of this
similarity, we call this
a \emph{role-flow approach}. An alternative way to describe the FSO protocols is to consider them as
\emph{guarded actions\/}~\cite{Dijkstra75} whose guards express the successful enrollment of agents.

An important aspect of FSO is the fact that the coordinator does not care whether a candidate actor
is a professional carer, a volunteer, a patient, or a machine.
The agent just needs to \emph{qualify\/} for the sought role. This means that agents need to be known
beforehand in order to play significant roles in an FSO.
Knowing an agent means having a semantic description of what the agent may do. More information
about this ma be found in~\cite{de2014mutualistic,SDB13a}. No semantic processing is used
in our simulation models due to their simple formulation.

If the coordinator can find agents for all the roles it requires, the protocol starts. If not, there is what we
call a \emph{role exception}. This
means that the circle coordinator publishes a new event in the ``parent circle''-- the circle that includes
the current circle as a node. If not all the requested agents can be enrolled in the parent circle, this results
in a new role exception. Through this mechanism the coordinators ``spread the news'' about the
missing roles throughout the levels of the network.
As soon as enrollment is completed, the enrolled agents become a new transient FSO whose aim and
lifespan is defined by the execution of their associated protocol.
We call said transient FSO a ``social overlay network.''

This realizes inter-organizational collaboration and shifts the responsibility
for service response composition from the user to the ``system''---more information
about this is provided in~\cite{DeFPa15a}.

FSO circles are structured as so-called service-oriented communities (SoC's).
They are described in detail in~\cite{DeBl10,DFCBD12}.
A special case of SoC is given by a person and a periphery of devices that the person has access to.
We call such SoC an individual SoC (iSoC).

\begin{figure}[h]
        \centerline{\includegraphics[width=0.5\textwidth]{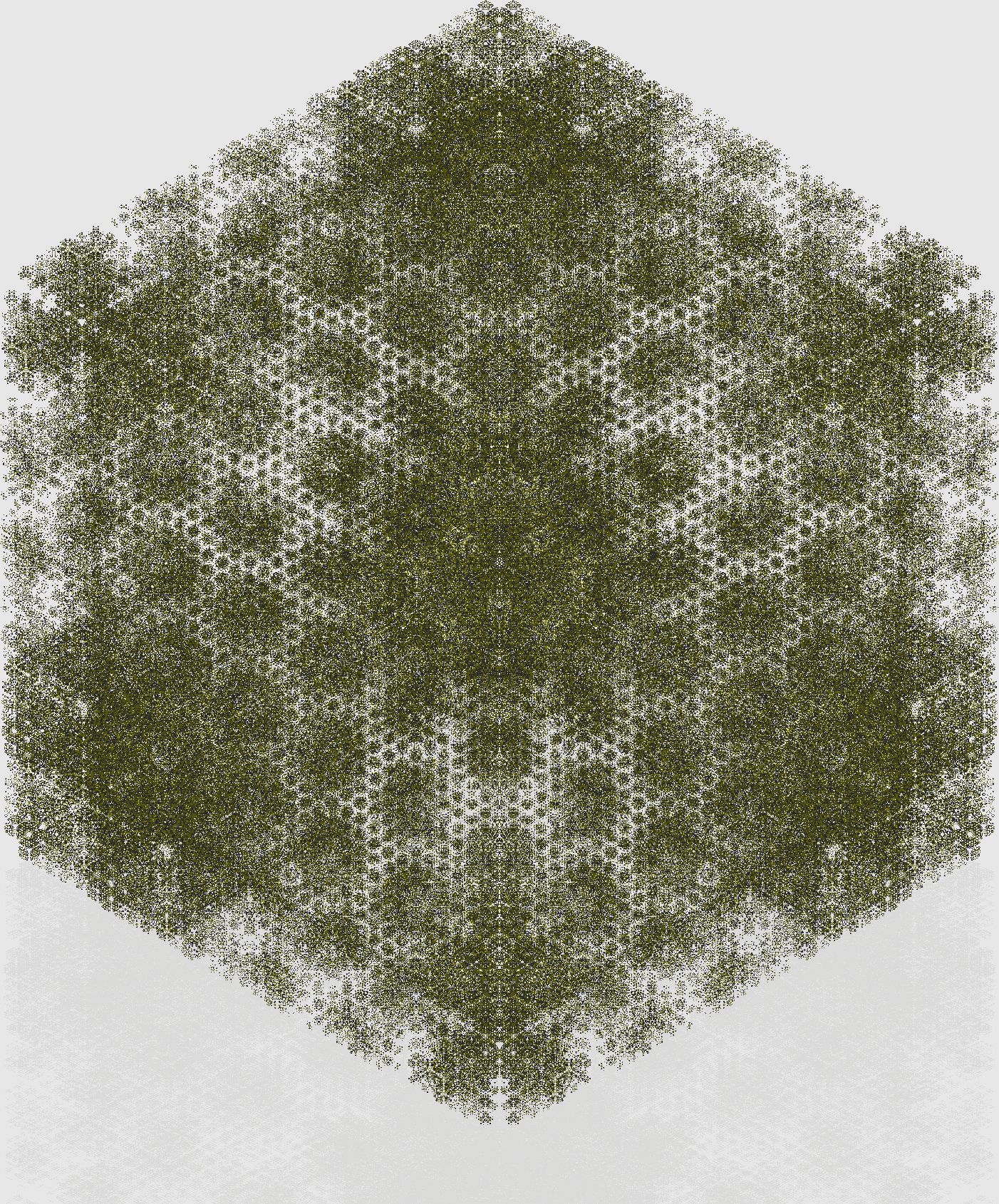}}
\caption{Tridimensional representation of all the social overlay networks derivable
from FSO ``000111234555''. Rendering is done via POV-Ray~\cite{povray}.}\label{f:fso}
\end{figure}

Figure~\ref{f:fso} represents the set of all possible social overlay networks in an FSO
including six roles (roles 0--5) and 12 agents distributed as follows:
role 0, role 1, and role 5 are played by 3 agents each while role 2, role 3, and role 4 are played by
1 agent each.

\section{Simulation Model}\label{s:act}

For our simulations
we use NetLogo~\cite{tisue_netlogo:simple_2004}, a tool that allows agent-based rapid
prototyping.
Netlogo provides a simulation area that functions as a virtual world in which events are triggered on the
initialized agents in discrete time steps called ``ticks''. The simulation area includes several classes of agents,
which are described in what follows.

\begin{Definition}[Elderly Agent]
Elderly agents \textup{(EA'}$\!$s\textup{)} are agents representing elderly or impaired persons. \textup{EA'}$\!$s are assumed not to leave
their house, in which their condition is monitored by Device Agents (see Definition~\ref{d:DA}).
\label{d:EA}
\end{Definition}

\begin{Definition}[Professional Carer]
\hfill Professional carers \textup{(PC'}$\!$s\textup{)} represent agents able to supply certified healthcare services
to other agents. Such agents are \emph{institutional}, meaning that their action is regulated by
rules, laws, and a professional code of service. This translates into relative high availability and
guarantees of timely intervention.
\end{Definition}

\begin{Definition}[Informal Carer; Verification]
\hfill In- formal carers \textup{(IC'}$\!$s\textup{)} are mobile agents that are able to provide non-professional services.
In what follows it is assumed that \textup{IC'}$\!$s can move to the location of an \textup{EA} and report
whether that \emph{EA} has truly experienced a fall or not.
This operation is called in what follows \emph{verification}. Symbol $V(x)$ shall be used
to represent a verification carried out by agent $x$.
\end{Definition}

\begin{Definition}[Device Agent]\label{d:DA}
Device agents \textup{(DA'}$\!$s\textup{)} are simple purposefully-behaviored~\textup{\cite{RWB43}} devices
(as, e.g., accelerometers, gyroscopes,
motion sensors, cameras, combustion detectors, etc.) that are able to provide domotica or telemonitoring services.
In particular an accelerometer and a gyroscope \textup{DA} both trigger an alarm when they ascertain, with a certain probability,
that an \textup{EA} has fallen. Another example of \textup{DA} is a motion sensor estimating the motion vector
of an \textup{EA} moving in his or her house.
\end{Definition}
As already mentioned, DA are characterized by a non-zero probability of FP's
and FN's. Thus for instance in some cases an accelerometer may detect a fall when
this is not true, while in some other cases that accelerometer may not detect a true case of fall.

\begin{figure}[h]
        \centerline{\includegraphics[width=0.5\textwidth]{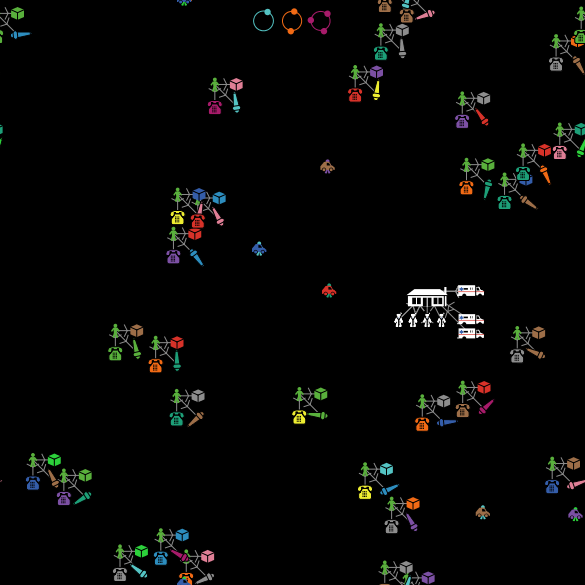}}
\caption{NetLogo screenshot of the simulation area with initialized agents.}\label{f:simarea}
\end{figure}

\begin{Definition}[Mobility Agent]
Mobile agents \textup{(MA'}$\!$s\textup{)} are agents able to provide extended mobility services to other agents. In what follows
a single type of \textup{MA} is employed, which represents ambulances.
\end{Definition}

\begin{Definition}[Community Agent; Canon]
\hfill Com- munity agents \textup{(CA'}$\!$s\textup{)} are agents coordinating a ``circle'' of other agents.
In what follows it is assumed
that \textup{CA}'s implement the set of rules and actions of \textup{SoC}'s, namely the building blocks
of the organizational structure introduced in Sect.~\ref{s:fso}.
The \textup{CA} rules and coordination actions are called the \emph{canon\/} of the \textup{FSO}~\textup{\cite{Koe67}}.
\textup{CA}'s may be implemented as middleware components
as described in, e.g., \textup{\cite{DF13c}}.
\label{d:CA}
\end{Definition}

\subsection{Simulation FSO}
The FSO adopted in our simulations is structured into three concentric levels:
\begin{enumerate}
\item The innermost level is constituted by iSoC's (see Sect.~\ref{s:fso}) coupling
      an EA with one or more DA's that continuously ``watch'' the EA in order to assess whether
      a fall event has taken place. Each of the iSoC also includes a CA, called in what follows
      ``L1-CA'' (level-1 CA).
      iSoC's correspond to primary users of ambient assisted living (AAL) services~\cite{aalstake}.
      They could be visualized, e.g., as a smart house inhabited by an elderly person.
\item The middle level is an SoC that includes the iSoC's (represented by their L1-CA's), a number
      of IC agents, and a coordinating CA. The latter shall be called in what follows ``L2-CA'' (level-2 CA).
      This SoC corresponds to primary and secondary users (\emph{sensu}~\cite{aalstake}).
      The elderly people, their relatives, and a ``cloud'' of informal carers may be used to visualize
      the second level of our FSO.
\item The outermost FSO level is constituted by the middle level SoC (represented by
      its L2-CA), a number of PC and MA agents, and a coordinating CA (``L3-CA'').
      This SoC represents all the AAL stakeholders in a certain region---including
      the infrastructures those stakeholder have access to.
\end{enumerate}

\begin{figure}[ht!]
        \centerline{\includegraphics[width=0.5\textwidth]{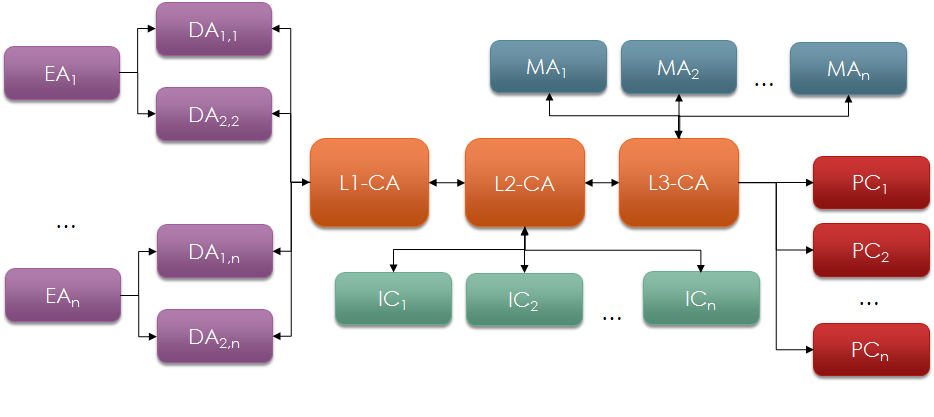}}
\caption{A representation of the FSO employed in our simulation models.}\label{f:elderpersonsFSO}
\end{figure}

As already discussed in Sect.~\ref{s:fso},
the agents associated with a coordinating CA send it notifications.
Whenever there is a need for communication between the CA's of distinct levels, ``exception'' messages are
triggered. It is assumed that these messages are transmitted reliably and instantaneously.
The overall FSO structure is depicted in Fig.~\ref{f:elderpersonsFSO}.

The operation of the FSO is as follows:
the EA's in the innermost iSoC may or may not experience falls. At the same time, the
DA's associated to the EA's may or may not issue their ``fall detected'' event.
When they do, the event reaches L1-CA. 

Four are the major protocols that may be executed:
\begin{enumerate}
\item Protocol $p_1$ corresponds to a request
for care requiring a PC and a MA. As both roles are missing, this
triggers an exception, namely an event for L2-CA. No agents may be enrolled in the middle level of
the FSO as that SoC lacks PC and MA agents.
As a consequence, a new exception ensues. The event thus reaches L3-CA where agents are enrolled
and the readied protocol $p1$ is finally launched.
Because of this, a general practitioner leaves his or her hospital
with an ambulance directed to the house the ``fail detected'' event originated from.
\item Protocol $p_2$ corresponds to a request for verification.
As verification requires an IC agent it may be considered as
a secondary care intervention. In this case the event is resolved in the middle level
of the FSO. Enrollment follows and $p_2$ is launched.
The verification step is concluded with either a confirmation
or a cancellation
of the ``fall event''. In the latter case, a ``FP event'' is issued by the IC agent.

\item As we have just described,
an ``FP event'' takes place in the middle level of the FSO and readies a third protocol,
$p_3$.
Said protocol is only used to forward the FP event to the outermost level. To do so,
the protocol simply ``enrolls'' the L3-CA agent. Obviously L2-CA has no actor able
to play role L3-CA, thus an exception is raised and delivers the FP event to its intended consignee.
\item The arrival of the FP event at L3-CA
 triggers a fourth protocol, $p_4$, which requires no roles and thus immediately cancels $p_1$.
\end{enumerate}

Canceling a protocol means that, \emph{if the protocol is still being executed,}
its execution is aborted. Actors involved in a canceled protocol are again available
for enrollment in readied protocols. In the case of a PC and MA, canceling a $p_1$ protocol
makes both agents reach their ``base of operation'' (i.e., a hospital.)

Note that if the execution of $p_1$ is very fast, the PC and MA may reach the EA sooner than the IC.
In a such a case it is $p_2$ that is canceled. If so, the enrolled IC is ``freed'' and resumes
its ``random walk'' through the virtual world.

A visualization of the simulation area containing
the initialized agents can be seen in
Fig.~\ref{f:simarea}. The green ``person'' shape represents the EA agents, while the connected devices represent
the DA's.
The white building represents the hospital, and connected to it we have the MA's and PC's.

\section{Simulation Scenarios}\label{s:sce}

Here we first introduce in Sect.~\ref{s:sce:met} a number of figures used to summarize the results of our simulations.
Simulation scenarios are then described in Sect.~\ref{s:sce:sce}.

\subsection{Evaluation metrics}\label{s:sce:met}
\begin{Definition}[Social Cost]
We shall call social cost (\textit{SC}) of agent $A$ the number of cycles that $A$ makes use of to intervene
for either a true alarm or a FP.
\end{Definition}
From its definition it is apparent that, in case of a FP, the social costs associated with
the agents responding to the alarm are wasted.

\begin{Definition}[Cumulative Social Cost (\totalcost)]
Metric \totalcost{} is defined as the overall number of
cycles used by all of a community's agents to deal with
the true alarms and FP's experienced during a simulation run.
\end{Definition}

\begin{Definition}[Number of Treated Cases]
\hfill The number of treated cases (\,\NTC) is the total amount
of alarms---either true cases or \textup{FP}---that took place
in the course of a simulation run.
\end{Definition}

\begin{Definition}[Normalized \totalcost]
Normalized cumulative social cost (\,\NCSC)
is given by the ratio \( \frac{\totalcost}{\NTC}. \)
\end{Definition}

\begin{Definition}[Waiting Time] Waiting time \textup{(WT)} is the number of cycles elapsed from the event of a triggered alarm
to either of the following events:
the arrival of care; the canceling of the care request; confirmation of FP; the end of the simulation run.
\end{Definition}

\begin{Definition}[Cumulative Waiting Time (\totaltime)] \totaltime{} is the
sum of all the individual \textup{WT'}$\!$s that took place during a simulation run.
\end{Definition}

\begin{Definition}[Normalized \totaltime]
Metric \hfill \NTT{} is given by the ratio
is given by the ratio \( \frac{\totaltime}{\NTC}. \)
\end{Definition}

\subsection{Scenarios}\label{s:sce:sce}

In our experiments we distinguish two classes of scenarios:
\begin{description}
\item[$S_1(X)$] In this scenario every EA is ``guarded'' by a single sensor to monitor for the occurrence
                of the fall event.  $X$ IC's are defined. When $X=0$, there is no level-2 FSO. 
Scenario $S_1$ is described through the pseudo-code in Table~\ref{t:S1}.

\item[$S_2(X)$] In this scenario two DA's are used per each EA (as suggested, e.g., in~\cite{EDC2013}).
In what follows we shall refer to the two DA's as an accelerometer ($A$) and a gyroscope ($G$).
$X$ is as in $S_1(X)$.
Scenario $S_2$ is described through the pseudo-code in Table~\ref{t:S2}.
\end{description}

As seen in Table~\ref{t:S1},
the AlarmEvent function keeps track of and returns either the number of ticks necessary to reach the
corresponding $EA$ or the number of ticks until a protocol $p_4$ is executed and the AlarmEvent
is interrupted.


\begin{table}[ht!]
\begin{small}
\begin{tabbing}
\hspace*{3ex}\=\hspace*{3ex}\=\hspace*{3ex}\=\hspace*{3ex}\=\hspace*{3ex}\=\hspace*{3ex}\=\hspace*{3ex}\=\hspace*{3ex}\kill
\textbf{Scenario} $S_1$\>\>\>\>\> // \textsf{A single DA, zero or more IC agents.}\\
\textbf{Begin}\\
\> // \totalfp:  total amount of \textit{FP} events\\
\> // \totalfn: total amount of \textit{FN} events\\
\> // \totalcost$(S_1)$: amount of ticks spent\ldots\\
\> // \hskip 32pt \ldots to manage a ``fall event'' (see Sect.~\ref{s:act})\\
\> // \pf: probability to experience a fall\\
\> // \pfn: probability of a \textit{FN} event\\
\> // \pfp: probability of a \textit{FP} event\\[6pt]
\> $\totalfp \leftarrow 0,  \totalfn \leftarrow 0, \totalcost(S_1) \leftarrow 0, \NTC \leftarrow 0$;\\[4pt]
\> \textbf{For all} ticks $\in T$ \textbf{Do}\\
\>\> // Toss(\pf) returns 1 with probability \pf,\ldots \\
\>\> // \ldots and 0 with probability 1 - \pf\\[4pt]
\>\> TrueFall $\leftarrow$ Toss(\pf);\\
\>\> \textbf{If} (TrueFall $\equiv$ 1) \textbf{Then}\\
\>\>\> // \emph{res} is 1 with probability \pfn\\
\>\>\> \emph{res} $\leftarrow$ Toss(\pfn);\\
\>\>\> \textbf{If} (\emph{res} $\equiv$ 1) \textbf{Then}\\
\>\>\>\> // (There is a fall) $\wedge$ (fall goes undetected)\\
\>\>\>\> \emph{FN} $\leftarrow 1$;\\
\>\>\> \textbf{Else}\\
\>\>\>\> // (There is a fall) $\wedge$ (fall is correctly detected)\\
\>\>\>\> // AlarmEvent returns the number of ticks \ldots\\
\>\>\>\> // \ldots necessary to reach the fallen EA\\
\>\>\>\> \emph{cost} $\leftarrow$ AlarmEvent()\\
\>\>\>\> // Number of treated cases is updated\\
\>\>\>\> $\NTC \leftarrow \NTC + 1;$\\
\>\>\> \textbf{Endif}\\[4pt]
\>\> \textbf{Else} \hskip5pt // TrueFall is not set\\
\>\>\> // \emph{res} is 1 with probability \pfp\\
\>\>\> \emph{res} $\leftarrow$ Toss(\pfp);\\
\>\>\> \textbf{If} (\emph{res} $\equiv$ 1) \textbf{Then}\\
\>\>\>\> // (There is no fall fall) $\wedge$ (phantom fall is detected)\\
\>\>\>\> \emph{FP} $\leftarrow 1$;\\
\>\>\>\> // AlarmEvent returns the number of ticks\ldots \\
\>\>\>\> // \ldots until the EA is reached\ldots\\
\>\>\>\> // \ldots or until the service is interrupted\ldots\\
\>\>\>\> // \ldots by a call to protocol $p_4$.\\
\>\>\>\> \emph{cost} $\leftarrow$ AlarmEvent()\\
\>\>\>\> // Number of treated cases is updated\\
\>\>\>\> $\NTC \leftarrow \NTC + 1;$\\
\>\>\> \textbf{Else}\\
\>\>\>\> // (There is no fall) $\wedge$ (no phantom fall is detected)\\
\>\>\>\> // Thus, do nothing! \\
\>\>\> \textbf{Endif}\\
\>\> \textbf{Endif}\\[4pt]
\> \totalfn $\leftarrow$ \totalfn{} + \emph{FN};\\[4pt]
\> \totalfp $\leftarrow$ \totalfp{} + \emph{FP};\\[4pt]
\> \totalcost$(S_1)$ $\leftarrow$ \totalcost$(S_1)$ + \emph{cost};\\[4pt]
\> \textbf{Enddo}\\
\textbf{End}
\end{tabbing}
\end{small}
\caption{Pseudo-code of Scenario $S_1$.}
\label{t:S1}
\end{table}

In scenario $S_1$ and $S_2$ we consider the following configuration of agents:
\begin{itemize}
\item 30 EA's residing in their houses.
House locations are assigned in cells chosen pseudo-randomly
in the virtual world. The same distribution of cells is used in each run of the simulations.
\item 1 hospital (Level 3 CA), located in a cell chosen as described above.
\item 6 PC's and 5 MA's (ambulances) are located with the hospital.
\end{itemize}

We assume a certain probability of an actual ``fall event''. Let us call \pf{} said probability.
In what follows, $\pf=\frac1{600}$.

Once an actual fall takes place, we shall call \pfn{} the probability that that fall shall \emph{not\/}
be detected.
Likewise, every time a fall \emph{does not\/} take place, we shall call \pfp{} the probability that a fall is
(erroneously) detected.

It is important to highlight how
\pfn{} and \pfp{} are normally specific of a given sensor; to facilitate the description of
our experiments, in what follows we shall assume
that both $A$ and $G$ have the same \pfn{} and \pfp{}, respectively
equal to $\frac15$ and $\frac1{500}$. Corresponding events are assumed to be
independent of each other (in other words, there is no correlation between
those events).

\begin{table}[ht!]
\begin{small}
\begin{tabbing}
\hspace*{3ex}\=\hspace*{3ex}\=\hspace*{3ex}\=\hspace*{3ex}\=\hspace*{3ex}\=\hspace*{3ex}\=\hspace*{3ex}\=\hspace*{3ex}\kill
\textbf{Scenario} $S_2$\>\>\>\>\> // \textsf{Two DA's, zero or more IC agents.}\\
\textbf{Begin}\\
\> // \totalfp:  total amount of \textit{FP} events\\
\> // \totalfn: total amount of \textit{FN} events\\
\> // \totalcost$(S_2)$: amount of ticks spent\ldots\\
\> // \hskip32pt \ldots to manage a ``fall event'' (see Sect.~\ref{s:act})\\
\> // \pf: probability to experience a fall\\
\> // \pfn: probability of a \textit{FN} event for both $A$ and $G$\\
\> // \pfp: probability of a \textit{FP} event for either $A$ or $G$\\[6pt]
\> $\totalfp \leftarrow 0$,  $\totalfn \leftarrow 0$, $\totalcost(S_2) \leftarrow 0, \NTC \leftarrow 0$;\\[4pt]
\> \textbf{For all} ticks $\in T$ \textbf{Do}\\
\>\> // Toss(\pf) returns 1 with probability \pf,\ldots \\
\>\> // \ldots and 0 with probability 1 - \pf\\[4pt]
\>\> TrueFall $\leftarrow$ Toss(\pf);\\
\>\> \textbf{If} (TrueFall $\equiv$ 1) \textbf{Then}\\[4pt]
$A_1$:
\>\>\> \emph{res} $\leftarrow$ Toss(\pfn);\\
\>\>\> \textbf{If} (\emph{res} $\equiv$ 1) \textbf{Then}\\
\>\>\>\> // (There is a fall) $\wedge$ (fall is undetected by $A$)\\
\>\>\>\> $\FNA \leftarrow 1$;\\
\>\>\> \textbf{Else}\\
\>\>\>\> // (There is a fall) $\wedge$ (fall is correctly detected by $A$)\\
\>\>\>\> // AlarmEvent returns the number of ticks\ldots \\
\>\>\>\> // \ldots necessary to reach the fallen EA\\
\>\>\>\> \emph{cost} $\leftarrow$ AlarmEvent();\\[2pt]
\>\>\>\> // Number of treated cases is updated\\
\>\>\>\> $\NTC \leftarrow \NTC + 1;$\\
\>\>\> \textbf{Endif}\\[4pt]

$G_1$:
\>\>\> \emph{res} $\leftarrow$ Toss(\pfn);\\
\>\>\> \textbf{If} (\emph{res} $\equiv$ 1) \textbf{Then}\\
\>\>\>\> // (There is a fall) $\wedge$ (fall is undetected by $G$)\\
\>\>\>\> $\FNG \leftarrow 1$;\\
\>\>\> \textbf{Else} \>\>  // (There is a fall) $\wedge$ (fall is correctly detected by $G$)\\
\>\>\>\> // If the fall was not already detected by $A$\ldots\\
\>\>\>\> \textbf{If} ($\FNA\not\equiv1$) \textbf{Then}\\
\>\>\>\>\> \emph{cost} $\leftarrow$ AlarmEvent(); \hskip7pt $\NTC \leftarrow \NTC + 1;$\\
\>\>\>\> \textbf{Endif}\\
\>\>\> \textbf{Endif}\\[4pt]
\>\>\> \emph{FN} $\leftarrow \FNA\wedge\FNG$;\\
\>\>\> ResetAllFlags();\\
\>\> \textbf{Else} \hskip5pt // TrueFall is not set\\
$A_2$:
\>\>\> \emph{res} $\leftarrow$ Toss(\pfp);\\
\>\>\> \textbf{If} (\emph{res} $\equiv$ 1) \textbf{Then}\\
\>\>\>\> // (There is no fall) $\wedge$ (phantom fall is detected by $A$)\\
\>\>\>\> $\FPA \leftarrow 1$;\\
\>\>\> \textbf{Endif}\\
$G_2$:
\>\>\> \emph{res} $\leftarrow$ Toss(\pfp);\\
\>\>\> \textbf{If} (\emph{res} $\equiv$ 1) \textbf{Then}\\
\>\>\>\> // (There is no fall) $\wedge$ (phantom fall is detected by $G$)\\
\>\>\>\> $\FPG\leftarrow 1$;\\
\>\>\> \textbf{Endif}\\
\>\>\> \emph{FP} $\leftarrow \FPA\vee\FPG$;\\[4pt]
\>\>\> \textbf{If} (\emph{FP} $\equiv$ 1) \textbf{Then}\\
\>\>\>\> // AlarmEvent returns the number of ticks\ldots \\
\>\>\>\> // \ldots until the EA is reached or until the\ldots\\
\>\>\>\> // \ldots service is interrupted by a call to $p_4$.\\
\>\>\>\> \emph{cost} $\leftarrow$ AlarmEvent(); \hskip7pt $\NTC \leftarrow \NTC + 1;$\\
\>\>\> \textbf{Endif}\\
\>\> \textbf{Endif}\\[4pt]
\> $\totalfn \leftarrow \totalfn + \hbox{\emph{FN}}$;
   $\totalfp \leftarrow \totalfp + \hbox{\emph{FP}}$;\\[4pt]
\> $\totalcost(S_2) \leftarrow \totalcost(S_2) + \hbox{\emph{cost}}$;\\[4pt]
\> ResetAllFlags();\\
\> \textbf{Enddo}\\
\textbf{End}
\end{tabbing}
\end{small}
\caption{Pseudo-code of Scenario $S_2$.}
\label{t:S2}
\end{table}

%

%

We now describe the results obtained in scenarios $S_1$, $S_2$, and $S_3$.

\section{Results}\label{s:res}

Experiments are run for $X=0$ to 40 by increasing $X$ by five IC's per experiment.
Each experiment lasts 10000 ``ticks''.
Results for the $S_1$ scenarios are shown in 
Table~\ref{t:s1} while for the $S_2$ scenarios they are shown in Table~\ref{t:s2}.

Let us first consider FP, FN, Sensitivity, and Specificity. If we compare scenario $S_1$ to $S_2$ 
without IC's we can observe that in $S_2$ the number of FP's increases by more than 100 FP units. This
is due to the addition of a second DA: as can be seen from Table~\ref{t:S2},
with two DA's the alarm is triggered as 
a result of an OR operation of the alarms of each DA~\cite{DeDe02c}.
Moreover, by increasing $X$ we can observe that FP events increase too. This is because the 
higher the number of IC's, the sooner validation will occur.

\begin{figure}[h!]
	\centerline{\includegraphics[width=0.5\textwidth]{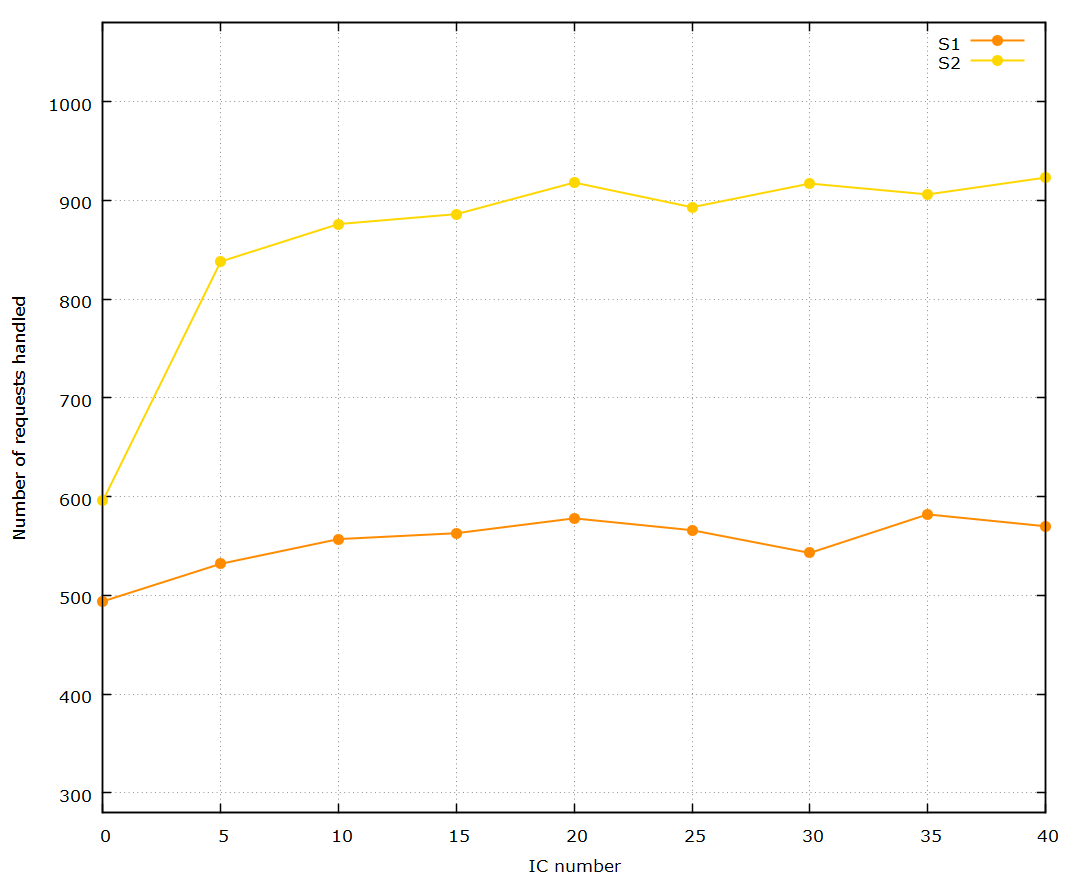}}
\caption{Number of alarm requests handled by the system for $S_1$ and $S_2$ with a number of IC's ranging from 0 to 40.}\label{f:ReqsHandled}
\end{figure}

\begin{figure}[t!]	\centerline{\includegraphics[width=0.5\textwidth]{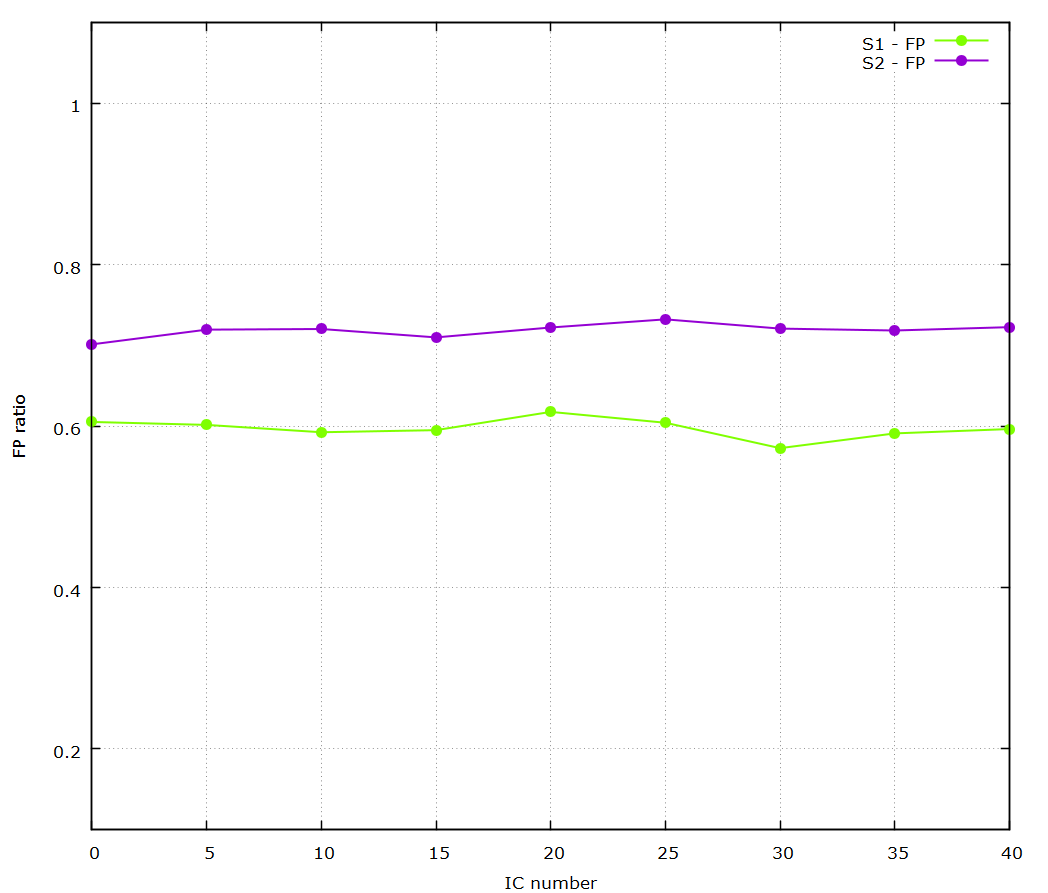}}
\caption{FP ratio for $S_1$ and $S_2$ with a
number of IC's ranging from 0 to 40.}\label{f:FPratio}
\end{figure}

\begin{figure}[b!]
\centerline{\includegraphics[width=0.5\textwidth]{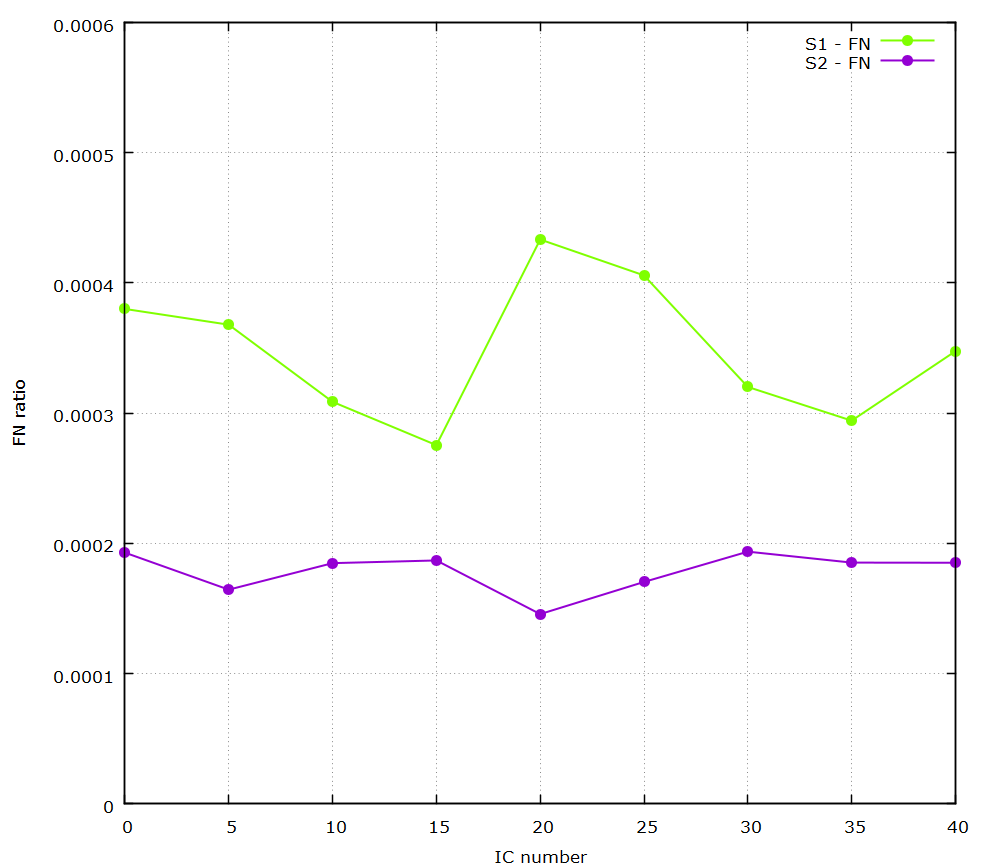}}
\caption{FN ratio for $S_1$ and $S_2$ with a
number of IC's ranging from 0 to 40.}\label{f:FNratio}
\end{figure}

\begin{figure}[t!]
	\centerline{\includegraphics[width=0.5\textwidth]{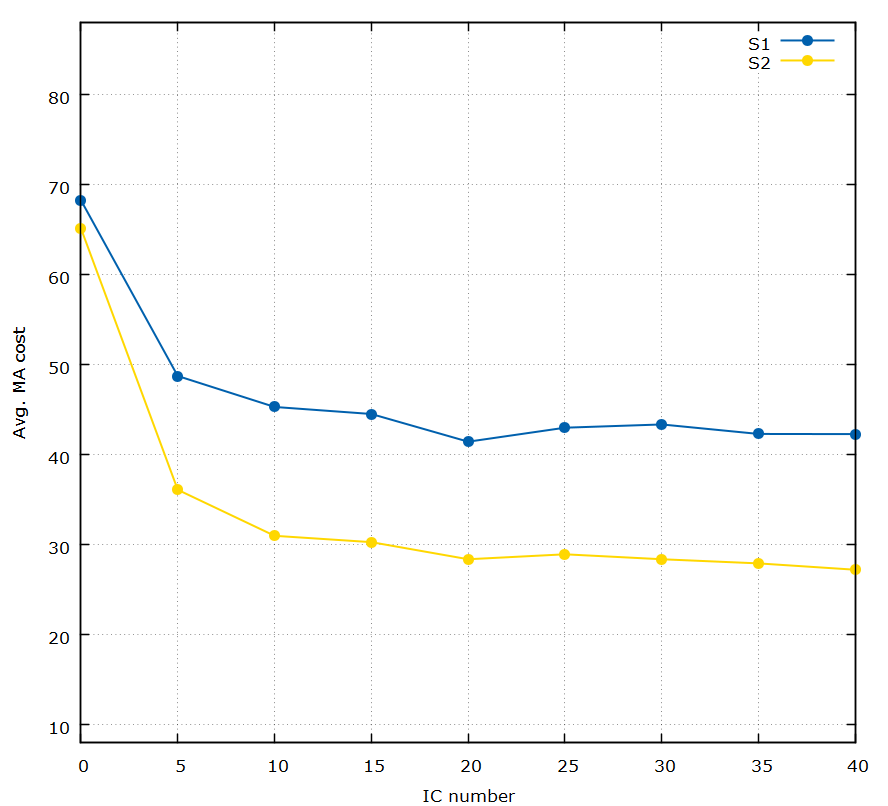}}
\caption{\NCSC{} (Average social costs) of $S_1$ and $S_2$ with a
number of IC's ranging from 0 to 40.}\label{f:avgMAcost}
\end{figure}

\begin{figure}[b!]
	\centerline{\includegraphics[width=0.5\textwidth]{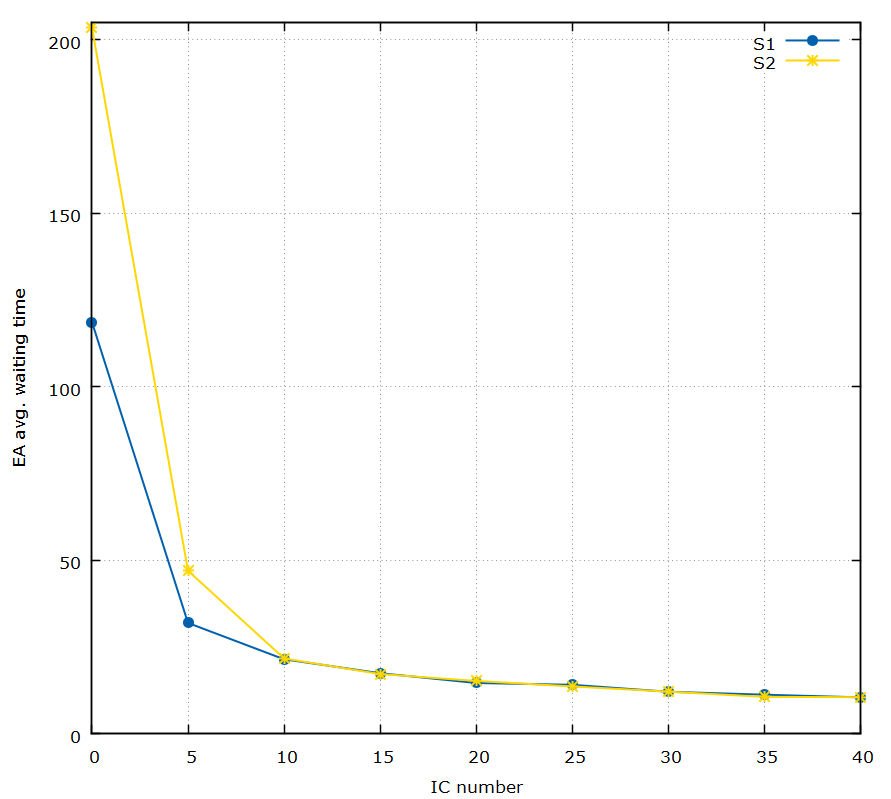}}
\caption{\NTT (Average waiting times) with  $S_1$ and $S_2$ and
a number of IC's ranging from 0 to 40.}\label{f:avgWtimes}
\end{figure}

Results can be seen in Figures~\ref{f:ReqsHandled}--\ref{f:avgWtimes}.

The FN values in $S_1$ are almost thrice those in $S_2$ for $X=0$.
This is because in $S_2$ an AND operation is performed 
between the FN outcome of DA1 and DA2 (see again Table~\ref{t:S2}).
 As a result the probability of having a FN decreases in $S_2$, and only 
when we have a FN from both alarms the FN is reported. With $X>0$
we see that the addition of IC's doesn't produce improvements. This is because in our simulation models
IC's cannot identify FN's. The 
graph in Figure~\ref{f:FPratio} depicts the FP ratios of $S_1$, and $S_2$ with various number of IC's. The FN 
ratios are visualized in Figure~\ref{f:FNratio}.

A larger number of alarms and a smaller number of FN's translates into a more sensitive system. This is
because by 
triggering more alarms, more cases where the alarm is truly positive are covered. On the other hand, there is no 
difference observed with regard to the specificity parameter for each scenario. From the results in 
Table~\ref{t:s1} and Table~\ref{t:s2} we see that in $S_1$, with or without volunteers, the average sensitivity is 
79.55\%, while for $S_2$ it increases to an average of 89.92\%. The Specificity values stay the same for all 
scenarios with an average close to 99\%.

We now focus our attention to social costs.
If we compare $S_1(0)$ with the case where IC's are added, we observe a 
significant decrease in the cost of MA's. This is because the IC's help verifying the actual condition
of the EA agents, thus reducing the cost of MA's. The same applies 
for the $S_2$ scenarios. Figure~\ref{f:avgMAcost} depicts the relation between \NCSC{} and $X$ for 
both scenarios. In the case of $S_1$, the minimum is reached for 20 IC's after which average social cost appears to 
stabilize. Another metric that shows the impact of IC's in the MA cost is the number of alarm verification's. In 
all scenarios where IC's are present the number of verifications performed by means of MA's is reduced drastically.

A significant metric is the average time until a triggered alarm for an EA gets verified,
namely \NTT.
The average waiting time reflects the reaction time of the system for a given event. The best results are 
obtained with 10 IC's. Additional effort beyond this value produces little improvement. With 10 IC's
no significant differences are observed between $S_1$ and $S_2$. A graph showing the average waiting time in 
function of the number of IC's is given in Fig.~\ref{f:avgWtimes}.

We can conclude that the involvement of IC agents in the fall identification system using the FSO model
helps reducing the social costs, while at the same time the average response time to fall events is reduced.
Moreover, it transpires that the involvement of IC agents is beneficial up to some threshold---possibly
in function of the dimensions of the virtual word and other simulation parameters.

More and more complex experiments shall follow. In particular we plan to explore 
the relation between the exhibited
amount of social cooperation and an organization's ability to deliver and sustain a given state of ``welfare''.

\begin{table*}[htbp]
  \centering
    \begin{tabular}{lrrrrrrrrr}
    \toprule
          & $S_1(0)$    & $S_1(5)$ & $S_1(10)$ & $S_1(15)$ & $S_1(20)$ & $S_1(25)$ & $S_1(30)$ & $S_1(35)$ & $S_1(40)$ \\
    \midrule
    FP & 299   & 320   & 330   & 335   & 357   & 342   & 311   & 344   & 340 \\
    FN & 56    & 63    & 51    & 46    & 74    & 68    & 52    & 48    & 58 \\
    TP & 195   & 212   & 227   & 228   & 221   & 224   & 232   & 238   & 230 \\
    TN & 147313 & 171175 & 165171 & 166988 & 170681 & 167515 & 162385 & 163097 & 166825 \\
    Avg. FP/tick & 0.0299 & 0.032 & 0.033 & 0.0335 & 0.0357 & 0.0342 & 0.0311 & 0.0344 & 0.034 \\
    Avg. FN/tick & 0.0056 & 0.0063 & 0.0051 & 0.0046 & 0.0074 & 0.0068 & 0.0052 & 0.0048 & 0.0058 \\
    FP rate & 0.6052 & 0.6015 & 0.5924 & 0.595 & 0.6176 & 0.6042 & 0.5727 & 0.591 & 0.5964 \\
    FN rate & 0.00038 & 0.00037 & 0.00031 & 0.00027& 0.00043& 0.0004& 0.00032& 0.00029& 0.00035 \\
    Sensivity & 77.68 & 77.09 & 81.65 & 83.21 & 74.91 & 76.71 & 81.69 & 83.21 & 79.86 \\
    Specificity & 99.79 & 99.81 & 99.8  & 99.79 & 99.79 & 99.79 & 99.8  & 99.78 & 99.79 \\
    \totalcost (MA) & 33720 & 25928 & 25230 & 25062 & 23951 & 24333 & 23541 & 24619 & 24102 \\
    \totalcost (IC) & 0     & 0     & 11351 & 9252  & 7874  & 7414  & 6023  & 5963  & 5400 \\
    \totaltime   & 58617 & 16981 & 11943 & 9815  & 8451  & 7980  & 6565  & 6545  & 5969 \\
    \NTC& 494   & 532   & 557   & 563   & 578   & 566   & 543   & 582   & 570 \\
    $V$(IC) & 0     & 472   & 540   & 555   & 576   & 566   & 540   & 582   & 568 \\
    $V$(MA & 296   & 33    & 8     & 6     & 0     & 0     & 1     & 0     & 0 \\
    $I$(MA) & 193   & 211   & 225   & 227   & 220   & 222   & 231   & 235   & 230 \\
    \NCSC (MA) & 68.259& 48.737 & 45.296& 44.515& 41.438 & 42.991& 43.353& 42.301 & 42.284\\
    \NTT & 118.658 & 31.919& 21.442 & 17.433& 14.621& 14.099 & 12.09& 11.246 & 10.472 \\
    \bottomrule
    \end{tabular}%
  \caption{Results of $S_1$ scenarios. $V$ stands for ``verifications'', $I$ for ``interventions''.}
  \label{t:s1}%
\end{table*}%


\begin{table*}[htbp]
  \centering
    \begin{tabular}{lrrrrrrrrr}
    \toprule
          & $S_2(0)$    & $S_2(5)$ & $S_2(10)$ & $S_2(15)$ & $S_2(20)$ & $S_2(25)$ & $S_2(30)$ & $S_2(35)$ & $S_2(40)$ \\
    \midrule
    FP& 418   & 603   & 631   & 629   & 663   & 654   & 661   & 651   & 667 \\
    FN& 20    & 24    & 29    & 29    & 23    & 28    & 31    & 30    & 30 \\
    TP& 178   & 235   & 245   & 257   & 255   & 239   & 256   & 255   & 256 \\
    TN& 103699 & 145952 & 156918 & 155168 & 157907 & 164355 & 160108 & 161965 & 162078 \\
    Avg. FP/tick & 0.0418 & 0.0604 & 0.0631 & 0.0629 & 0.0663 & 0.0654 & 0.0661 & 0.0651 & 0.0667 \\
    Avg. FN/tick & 0.002 & 0.0024 & 0.0029 & 0.0029 & 0.0023 & 0.0028 & 0.0031 & 0.003 & 0.003 \\
    FP rate & 0.7013 & 0.7195 & 0.7203 & 0.7099 & 0.7222 & 0.7323 & 0.7208 & 0.7185 & 0.7226 \\
    FN rate & 0.00019& 0.00016& 0.00018& 0.00019 & 0.00015 & 0.00017& 0.00019& 0.00018& 0.00018\\
    Sensivity & 89.89 & 90.73 & 89.41 & 89.86 & 91.72 & 89.51 & 89.19 & 89.47 & 89.51 \\
    Specificity & 99.59 & 99.58 & 99.59 & 99.59 & 99.58 & 99.6  & 99.58 & 99.59 & 99.59 \\
    \totalcost (MA) & 38819 & 30249 & 27141 & 26829 & 26054 & 25832 & 26026 & 25296 & 25116 \\
    \totalcost (IC) & 0     & 33426 & 18114 & 14344 & 13048 & 11269 & 10199 & 8723  & 8785 \\
    \totaltime   & 121316 & 39364 & 18986 & 15229 & 13965 & 12160 & 11113 & 9628  & 9707 \\
    \NTC& 596   & 838   & 876   & 886   & 918   & 893   & 917   & 906   & 923 \\
    $V$(IC)& 0     & 745   & 824   & 877   & 913   & 887   & 913   & 904   & 922 \\
    $V$(MA)& 420   & 67    & 37    & 5     & 2     & 3     & 1     & 1     & 0 \\
    $I$(MA) & 161   & 219   & 228   & 235   & 237   & 226   & 239   & 232   & 232 \\
    \NCSC (MA) & 65.133 & 36.097 & 30.983 & 30.281& 28.381& 28.927& 28.382 & 27.921 & 27.211\\
    \NTT & 203.55& 46.974 & 21.673& 17.188& 15.212& 13.617& 12.119 & 10.627 & 10.517 \\
    \bottomrule
    \end{tabular}%
  \caption{Results of $S_2$ scenarios. $V$ stands for ``verifications'', $I$ for ``interventions''.}
  \label{t:s2}%
\end{table*}%



\section{Conclusions}\label{s:end}
\epigraph{We are greatly frustrated\\
by all our local, static organization\\
of an obsolete yesterday.}%
{{Richard Buckminster Fuller,}\\ \emph{Synergetics I}}

Traditional organizations are today confronted with unprecedented scenarios. As an example, the aging of the population, and even more the chronic diseases associated with aging place substantial demands on health and social care services. On the other hand, the ``old recipes'', which appeared to work in a less turbulent context, now reveal all their limitations. Traditional healthcare services, even when backed by the most advanced ICT, find it difficult to meet the ensuing ever increasing demand while guaranteeing both safety and cost-effectiveness. As observed by Buckminster Fuller, this is mainly due the inflexible and static organization of those services. As observed, e.g., in~\cite{Deloitte12}, the new context ``requires new ways of working.''

In this article we demonstrate one such way. By integrating the cost-effective context detection abilities of telecare devices with the superior cognitive functions of human beings we showed how it is possible to improve, to some extent and under certain conditions, the performance and quality of a health-critical service. Our preliminary results indicate that distributed and dynamic organizations able to tap into the wells of 
``social energy''~\cite{DF14d,DeBl10,de2014mutualistic} of our societies may constitute the basis for novel,
\emph{smarter solutions\/} to the new problems of our ever more complex world.

Our future experiments will 
aim at understanding how the size of the virtual world and the distribution of agents
influence the results. For this we are considering to design an ad hoc simulator
privileging performance to graphical rendering.

\bibliographystyle{abbrv}

\end{document}